\documentclass[twocolumn]{revtex4}
\usepackage{amsfonts}
\usepackage{amsmath}
\usepackage{amssymb}
\usepackage{hyperref}
\usepackage{ragged2e}
\usepackage{graphicx}
\usepackage[usenames,dvipsnames]{color}
\definecolor{darkblue}{RGB}{0,0,196}

\begin{document}

\title{Comparative Analysis of Jet and Underlying Event Properties Across Various Models as a Function of Charged Particle Multiplicity at 7 TeV\vspace{0.5cm}}

\author{Maryam Waqar$^{1}$, 
Haifa I. Alrebdi$^{2,}$\footnote{Corresponding author: hialrebdi@pnu.edu.sa},
Muhammad Waqas$^{3}$, 
K.S. Al-mugren$^{2}$, 
Muhammad Ajaz$^{1,}$\footnote{Corresponding author: ajaz@awkum.edu.pk},
\vspace{0.25cm}}

\affiliation{$^1$Department of Physics, Abdul Wali Khan University Mardan, 23200 Mardan, Pakistan\\
$^2$Department of Physics, College of Science, Princess Nourah bint Abdulrahman University, P.O. Box 84428,
Riyadh 11671, Saudi Arabia\\
$^3$School of Nuclear Science and Technology, University of Chinese Academy of Sciences, Beijing 100049, China\\
}

\begin{abstract}
\noindent {\bf Abstract:} 
In this study, a comprehensive analysis of jets and underlying events as a function of charged particle multiplicity in proton-proton (pp) collisions at a center-of-mass energy of $\sqrt{s} = 7$ TeV is presented. Various Monte Carlo (MC) event generators, including Pythia8.308, EPOS1.99, EPOSLHC, EPOS4$_{Hydro}$, and EPOS4$_{noHydro}$, are employed to predict particle production. The predictions from these models are compared with experimental data from the CMS collaboration. The charged particles are categorized into those associated with underlying events and those linked to jets. The analysis is restricted to charged particles with $|\eta| < 2.4$ and $p_{T} > 0.25$ GeV/c. Upon comparing the MC predictions with CMS data, it is observed that EPOS$4_{Hydro}$, EPOSLHC, and Pythia8 consistently reproduce the experimental results for all charged particles, underlying events, intrajet, and leading charged particles.
For charged jet rates with $p_{T}^{ch.jet} > 5$ GeV/c, EPOS4$_{Hydro}$ and Pythia8 perform exceptionally well. In the case of charged jet rates with $p_{T}^{ch.jet} > 30$ GeV/c, EPOSLHC reproduces satisfactorily good results, while EPOS4$_{Hydro}$ exhibits good agreement with the data at higher charged particle multiplicities as compared to the other models. This can be attributed to the conversion of energy into flow when "Hydro=on," leading to an increase in multiplicity.  EPOSLHC model described the data well due to the new collective flow effects, correlated flow treatment, and parametrization as compared to the EPOS1.99.
However, the examination of the jet $p_{T}$ spectrum and normalized charged $p_{T}$ density reveals that EPOS4$_{Hydro}$, EPOS4$_{noHydro}$, and EPOSLHC exhibit good agreement with the experimental results, while Pythia8 and EPOS1.99 do not perform as well due to the lack of correlated flow treatment.

\vspace{0.5cm}

\\

\end{abstract}

\maketitle

\section{Introduction}
The history of hadron production spans a broad and extensive scope in high-energy and nuclear physics. A comprehensive understanding of hadron and multiparticle production in hadron-hadron collisions remains an open question in the field of high-energy particle physics. At the energies attained in the Large Hadron Collider (LHC), proton-proton collisions predominantly result in inelastic interactions, giving rise to jets stemming from hard parton-parton scatterings with momentum exchanges on the order of several GeV/c. The soft interactions between partons and remnants account for the underlying event \cite{UE1,UE2}. Additionally, at low momentum transfer, diffractive processes and Multi-Parton Interactions (MPI) play pivotal roles in particle production. These partons originate from the strong interaction within hadrons \cite{cpc}. The theoretical modeling of particle production in such environments relies on Theoretical Models that are calibrated to match experimental data. In high-energy interactions, transfer momenta between partons occur at scales of many GeV/c, which are described by perturbative Quantum Chromodynamics (pQCD). Understanding particle production in proton-proton collisions at LHC energies necessitates a complete comprehension of the transition between the hard processes, governed by pQCD, and the soft processes, described by non-perturbative models of QCD. Jets are narrow, cone-shaped sprays of particles produced when high-energy quarks or gluons fragment and hadronize after being scattered in particle collisions.
Modern jet substructure techniques like grooming and the soft-drop algorithm offer sophisticated methods for studying the properties of jets in addition to the standard jet definitions used in this analysis. These methods are especially useful for improving the resolution of jet mass and other observables and minimizing contamination from soft, wide-angle radiation.
The soft-drop algorithm removes the softer, wide-angle components from the jet by applying specific criteria to the transverse momentum and to the angular separation of jet constituents. This process results in a cleaner jet structure, which is important for precisely identifying and measuring the properties of the originating particles, particularly in high-background environments like the Large Hadron Collider (LHC) \cite{jet1}. 
Enhancements to the original soft-drop technique, such as Recursive Soft Drop (RSD), have been made. RSD improves mass resolution and robustness against non-perturbative effects by repeatedly applying the soft-drop condition \cite{jet2}.The probability associated with the creation of a specific number of particles in a collision is referred to as multiplicity distributions \cite{P1, P2}. These distributions encapsulate all relevant information regarding particle correlations. In the context of Hadron-Hadron and Heavy Ion collisions, multiplicity distributions play a crucial role in understanding particle production mechanisms.
The mechanism governing particle production is linked to the probability \(p_{n}\) denoting the number of charged particles produced in the medium. The distributions of charged particles' multiplicities encompass detailed information about both soft and hard interactions. These multiplicity distributions stand as fundamental and ubiquitous observables in high-energy physics experiments. Moreover, they provide insight into various aspects of the particle production mechanism and the process of hadronization.
In this paper, we conduct a comprehensive analysis of jets and underlying events as a function of charged particle multiplicity in proton-proton collisions at $\sqrt{s}=7$ TeV. Various Monte Carlo models, namely Pythia8.3, EPOS4$_{Hydro}$, EPOS4$_{noHydro}$, EPOS1.99, and EPOSLHC, are employed for simulation, and their results are compared with CMS data. The simulation encompasses 1 million events. While our current study focuses on standard jet definitions, incorporating soft-drop and related grooming techniques in future analyses could potentially provide deeper insights and more precise measurements of the underlying event and jet properties.The structure of the remaining sections is outlined as follows: Section 2 delves into the Methods and Models utilized, Section 3 presents the Results and subsequent Discussion, and the Conclusion is provided in the final section.

\section{Models and method}
Pythia \cite{STM} is the most widely used event generator in high-energy physics and related areas. It can be used to simulate proton-proton collisions, as well as proton-antiproton and \(e^{+}e^{-}\) collisions. Pythia primarily simulates parton showers and the interactions between partons. Its ability to analyze Multi-Partonic Interactions \cite{MPI} and the Lund String Fragmentation Model \cite{LS1, LS2} is used for hadronization. Pythia simulates particle collisions through the following steps: hard scattering, parton showers, Initial State Radiation (ISR), Final State Radiation (FSR) \cite{ISR, FSR}, and finally, hadronization. Pythia employs the \(p_{T}\)-ordered approach \cite{2} for parton showers and uses the original impact parameter for multi-parton interactions \cite{3}. The Lund String Fragmentation model is used for hadronization \cite{4, 5}, which is the final step of fragmentation. For particle collisions, the energy of the particles must be greater than 10 GeV because, below this threshold, particles go into hadronic resonance, and Pythia fails to provide accurate results. Therefore, 10 GeV is chosen as the limit for the standard scale. In \(e^{+}e^{-}\) annihilation, this limit can be reduced, but in proton-proton collisions below this limit, the Pythia model is not reliable or trustworthy. Conversely, Pythia can be tested up to 100 TeV center-of-mass energy \cite{6, 6a,7, 7a}. 
The Pythia model is suitable for higher energy ranges. There is no internal facility for proton-nucleus and nucleus-nucleus collisions, but several programs interface with specific Pythia models, especially for decay processes and string fragmentation algorithms are available. Users must either use the HEPMC \cite{8} interface or write their own interface for simulation programs. Pythia events are always applicable at both the partonic and particle levels. Pythia8.3, written in C++, uses matching and merging techniques for parton showers and matrix elements.

EPOS is an event generator used for both cosmic ray air showers (EAS) simulations and heavy ion interactions. The high-energy hadronic interactions are described by the EPOS model, which includes parton remnants \cite{9}. EPOS is based on the string and quantum multiple scattering approach for various particle production mechanisms. It employs the Gribov-Regge-Parton-based Theory (GRPT) \cite{10} for soft interactions. The EPOS model also accounts for energy conservation at the amplitude level and centrality dependence in heavy ion collisions.

In EPOS1.99 \cite{epos1.99}, the data is tuned to Tevatron energies. EPOSLHC \cite{eposL} is an updated version of EPOS1.99, designed for Large Hadron Collider (LHC) energies \cite{ullah}. In EPOSLHC, different flow parametrizations for the core (small system but high-density matter) are introduced in proton-proton collisions compared to heavy ion collisions. The EPOSLHC model is tuned to 8 TeV, but some parameters are still missing for 13 TeV. In EPOSLHC, minimum bias results are reproduced for particles with transverse momentum distributions ranging from 0 to a few GeV/c \cite{13,14}. The EPOSLHC tune is also more accurate in reproducing multiplicity distributions at 7 TeV.

EPOS4 is an advanced Monte Carlo model framework \cite{epos4a, epos4b, R1} designed to simulate the full evolution of high-energy heavy ion collisions, including both initial state and final state interactions. In the EPOS4 approach, multiple scattering, either partonic or nucleonic, occurs in parallel, based on elementary considerations related to time scales. EPOS4 combines S-matrix theory (related to parallel scattering) with modern perturbative QCD approaches and saturation concepts. This parallel scattering approach distinguishes between "primary scattering" and "secondary scattering." In parallel scattering, the initial primary nucleon and their partonic constituents are involved, occurring instantaneously at very high energies. The S-matrix is a theoretical tool that uses a specific form of proton-proton scattering S-matrix (Gribov-Regge Theory) \cite{R1, R2, R3, R4}, which can also be used for nucleon-nucleon (AA) collisions. This feature offers a solid framework for understanding the initial dynamics of Quark-Gluon Plasma (QGP) and its hadronization. The EPOS4 model can mainly be used into two tunes: EPOS4\(_{Hydro}\) and EPOS4\(_{noHydro}\). In the EPOS4 with Hydro, full hydrodynamic evolution, hadronic cascade, core-corona procedure and equation of state are activated, while the other version operates without it. These two versions allow flexible simulations for different physical scenarios \cite{rinphys}. EPOSLHC, and EPOS4 are designed for LHC experiments and offer sophisticated features such as event-by-event fluctuations and complex initial state treatments. They are also flexible for cosmic ray simulations.

In our analysis, we utilized Rivet \cite{ri} to validate Monte Carlo event generators and compare the model predictions with the experimental data. RIVET has an extensive code library that compares event generator predictions with experimental data available on HEPData.

\section{Results and discussion}
We present a comprehensive analysis of jets and underlying event properties as a function of \( N_{ch} \) at \(\sqrt{s} = 7 \text{ TeV} \) in pp collisions. Different Monte Carlo models (Pythia8.308, EPOS1.99, EPOSLHC, and EPOS4) were used for simulations, and the MC predictions are compared with CMS data \cite{J}.

\subsection{Comparison with data}
\textbf{Jets and Underlying Event properties for charged particles:}

\begin{figure*}[p!]\
\begin{center}
\hskip-0.153cm
\includegraphics[width=8cm]{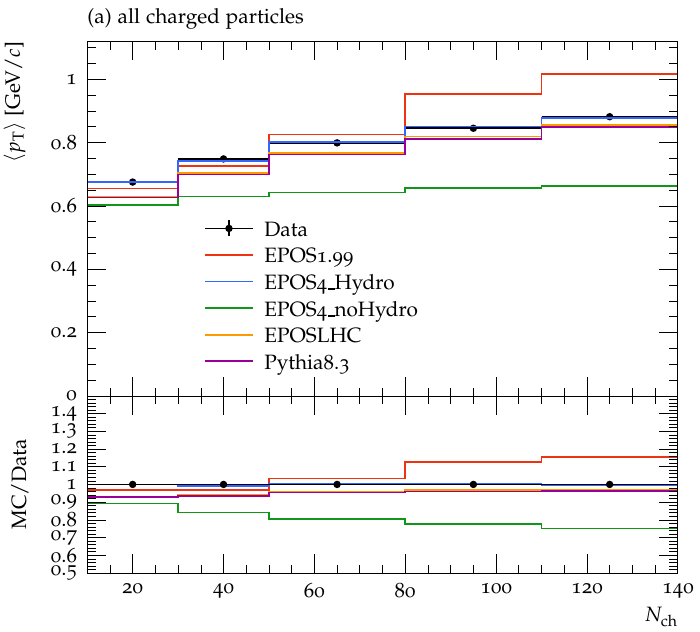}
\includegraphics[width=8cm]{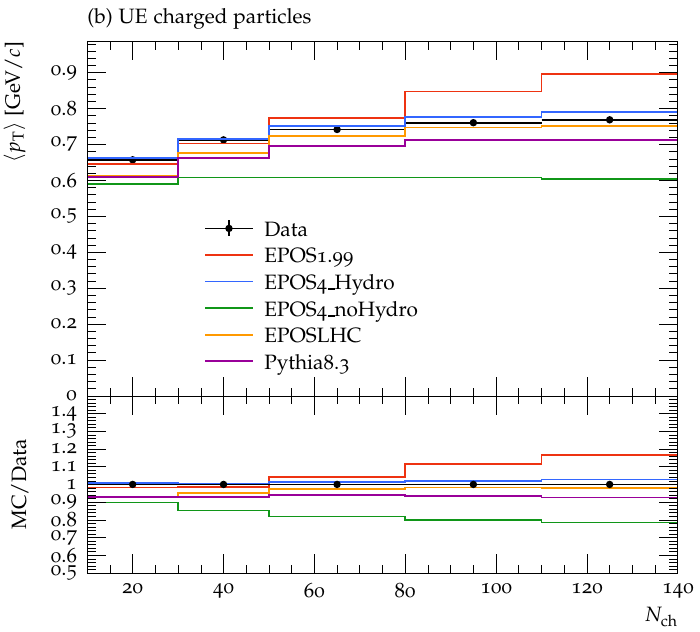}
\includegraphics[width=8cm]{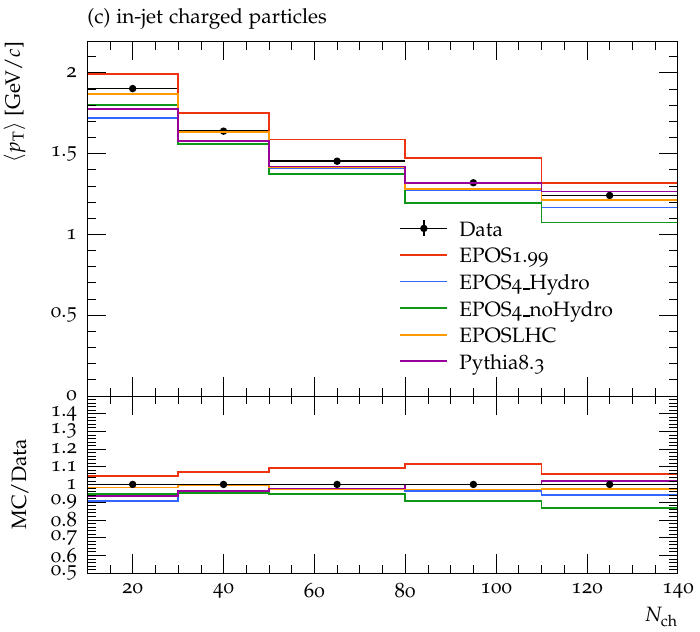}
\includegraphics[width=8cm]{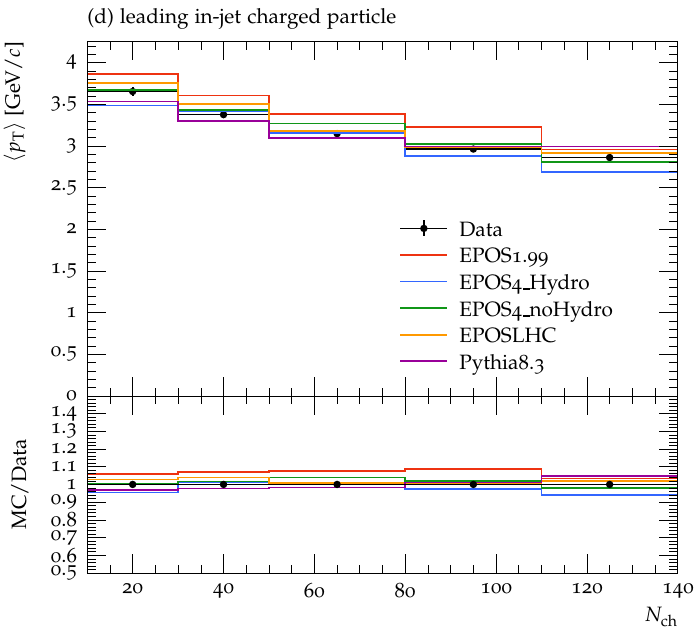}
\includegraphics[width=8cm]{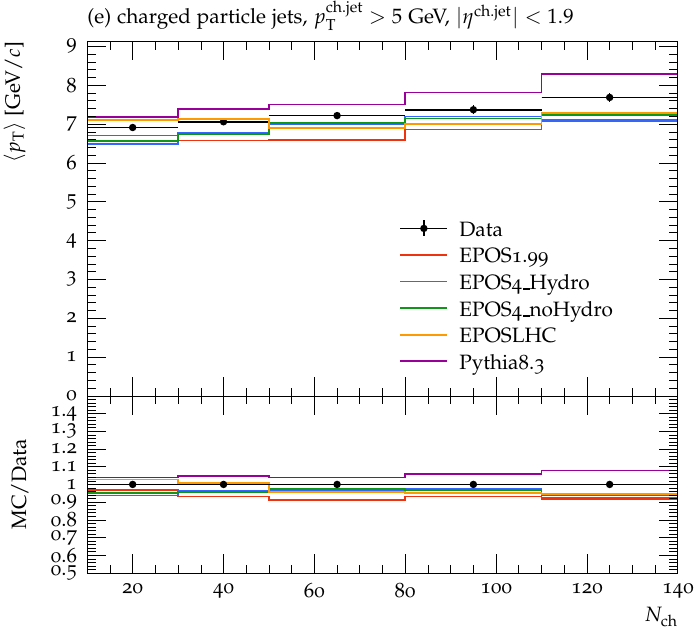}
\end{center}
\caption{Mean transverse momentum for (a) all charged particles (b) Underlying event(UE) charged particles (c) Intra-jet charged particles (d) Leading Intra-jet charged particles (e) charged particles jets as a function of charged particles multiplicity $N_{ch}$. Different Monte Carlo models, Pythia8.308\cite{STM}, EPOS1.99\cite{epos1.99}, EPOSLHC\cite{eposL}, EPOS4\cite{epos4a}, are compared with the experimental data \cite{J}}.\label{Fig1}
\end{figure*}

Figure~\ref{Fig1} illustrates the Mean transverse momentum ($<p_T>$) for (a) all charged particles (b) Underlying event (UE) charged particles (c) Intra-jet charged particles (d) Leading Intra-jet charged particles (e) charged particles jets as a function of charged particles multiplicity $N_{ch}$. The $<p_T>$ is increasing with an increase in charged multiplicity for all charged particles and underlying event (Fig. \ref{Fig1}(a \& b)) and decreasing with increasing charged multiplicity for jets (Fig. \ref{Fig1}(c, d \& e)). While comparing the Monte Carlo models with the data for all charged particles, all models show good agreement with the experimental data except EPOS1.99, which overpredicts for $N_{ch}>$ 80, and EPOS4$_{Hydro}$ underpredicts for $N_{ch}>$ 30. For UE-charged particles, once more, all models exhibit good predictions at low $N_{ch}$, while EPOS4$_{Hydro}$ and EPOSLHC accurately reproduce the data over the entire range. The Pythia8 model produces comparatively better predictions than EPOS1.99 and EPOS4$_{noHydro}$. EPOS1.99 overpredicts, while EPOS4$_{noHydro}$ underpredicts the data. 
For intra-jet and leading intra-jet charged particles, the models' comparison with the data show that all models accurately reproduce the data while EPOS1.99 over predicted and EPOS4$_{Hydro}$ underpredict at higher $N_{ch}>100$. This means that EPOS4$_{Hydro}$, EPOSLHC and Pythia8 models reproduce the data for the Underlying Event and intrajet charged particles for all the given $N_{ch}$. For Fig. \ref{Fig1}, we observed that EPOS4\(_{Hydro}\) accurately reproduces the results for higher \( N_{ch} \). This is because the hydro option increases the multiplicity and converts some energy into flow, resulting in the blue curve being much stronger than the other curves.
\begin{figure*}[htb!]
\begin{center}
\hskip-0.153cm
\includegraphics[width=8cm]{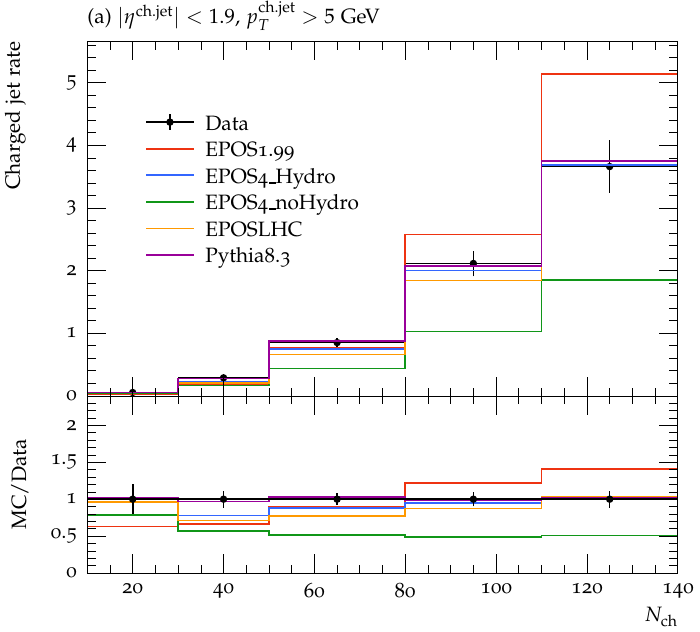}
\includegraphics[width=8cm]{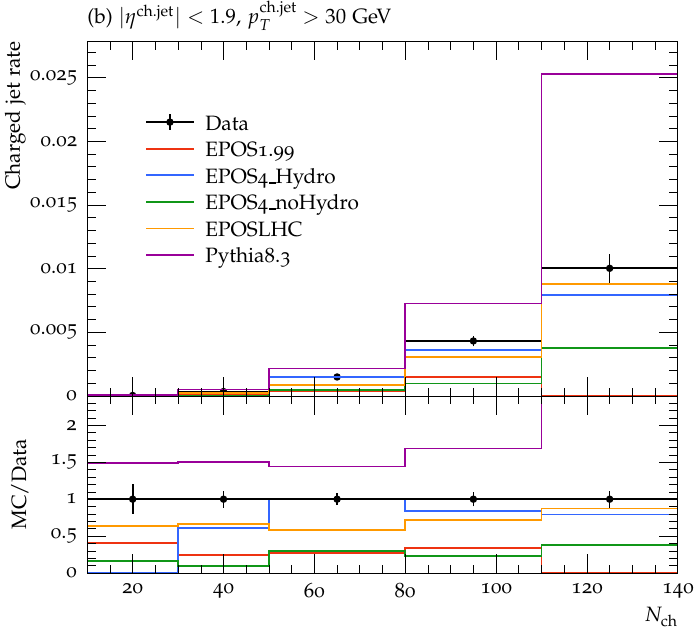}
\end{center}
\caption{Both panels show the number of charged particle jets vs. charged particle multiplicity for (a) \( p^{ch.jet}_{T} > 5 \) GeV/c and (b) \( p^{ch.jet}_{T} > 30 \) GeV/c in the region \(|\eta| < 1.9\). The prediction of different Monte Carlo models (Pythia8.308 \cite{STM}, EPOS1.99 \cite{epos1.99}, EPOSLHC \cite{eposL}, and EPOS4 \cite{epos4a}) are compared with the experimental data \cite{J}}.\label{Fig2}

\end{figure*}

\textbf{Jets properties for charged particles:}\\
In this section, we focus on the jet properties for charged particles, including the number of jets per event, differential jet \( p_T \) spectra, mean transverse momenta of jets, and jet widths, among other characteristics.
Figure~\ref{Fig2} illustrates the charged jet rate per event plotted as a function of \( N_{ch} \) for \( p^{ch.jet}_{T} > 5 \) GeV/c and \( p^{ch.jet}_{T} > 30 \) GeV/c.
The number of jets per event increases with the rising charged-particle multiplicity. This implies that a higher number of jets are generated for larger values of $N_{ch}$. The rates of jets increase from 0.05 jets/event to 4 jets/event as the charged particle multiplicity increases. In the case of $p^{ch.jet}_{T}>5GeV/c$, all models perform well in predicting the data, except the EPOS$4_{noHydro}$ model, which underestimates the results when $N_{ch}>20$. Models such as Pythia8, EPOS$4_{Hydro}$, EPOSLHC, and EPOS1.99 successfully reproduce the results for all given charged particle multiplicity distributions.

For $p^{ch.jet}_{T}>30GeV/c$, at low $N_{ch}$, only EPOSLHC effectively reproduces the results, while EPOS4$_{Hydro}$ disagrees with the data. At higher charged particle multiplicities, all models disagree except for the EPOS4\(_{Hydro}\) and EPOSLHC models, which show strong agreement with the data.
\begin{figure*}[p!]
\begin{center}
\hskip-0.153cm
\includegraphics[width=8cm]{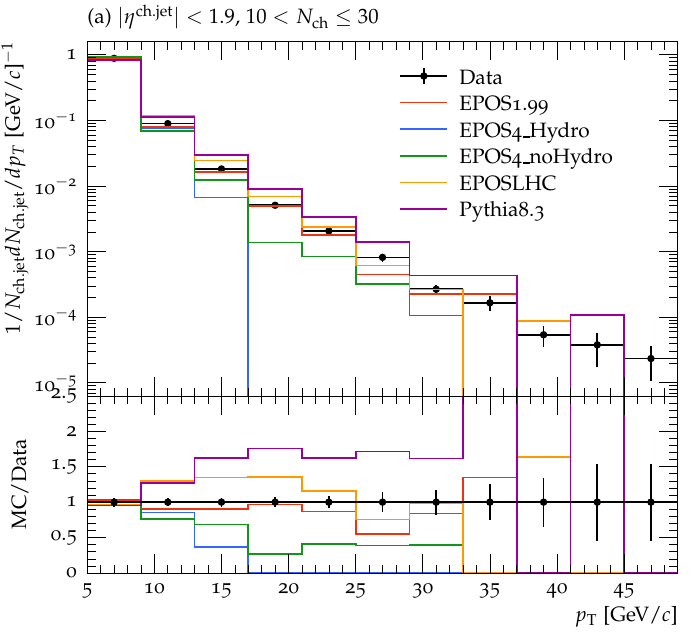}
\includegraphics[width=8cm]{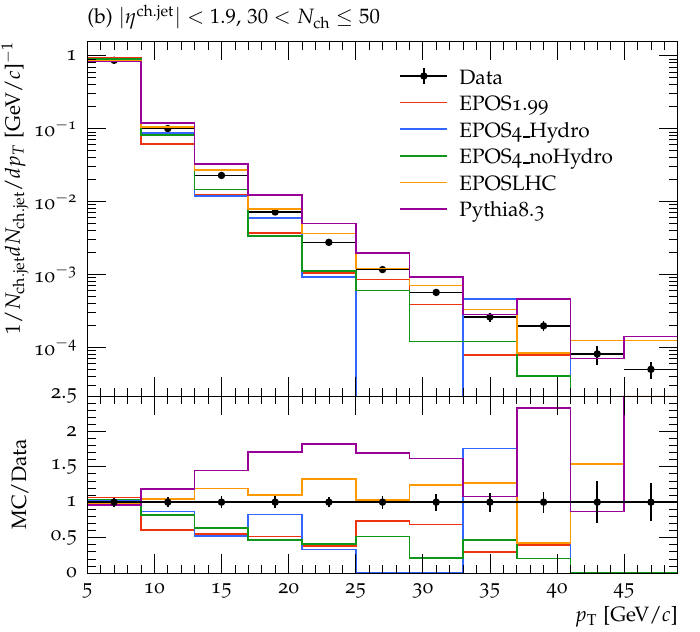}
\includegraphics[width=8cm]{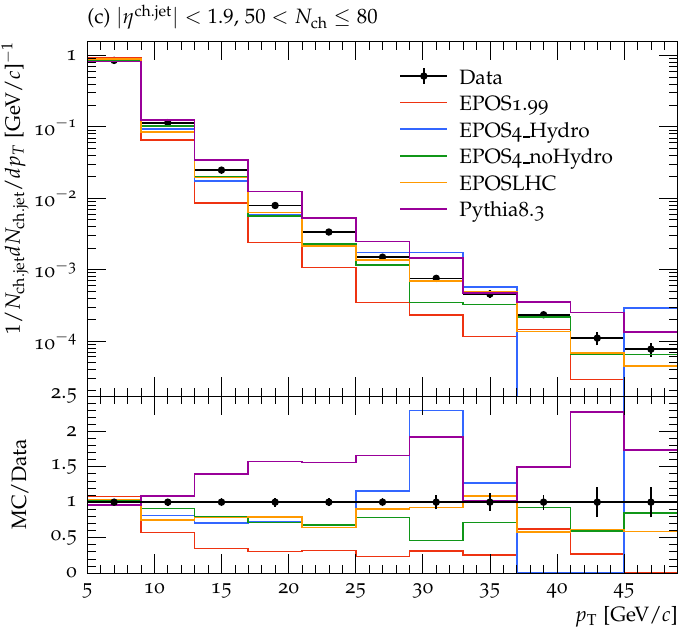}
\includegraphics[width=8cm]{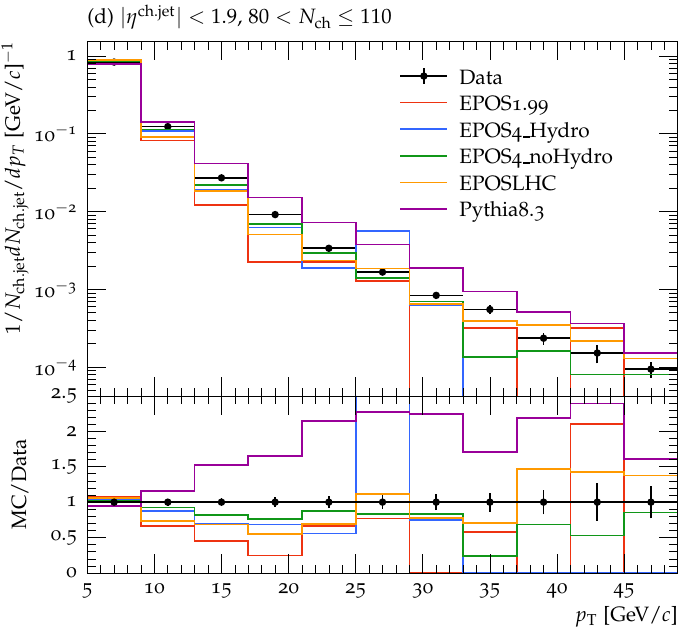}
\includegraphics[width=8cm]{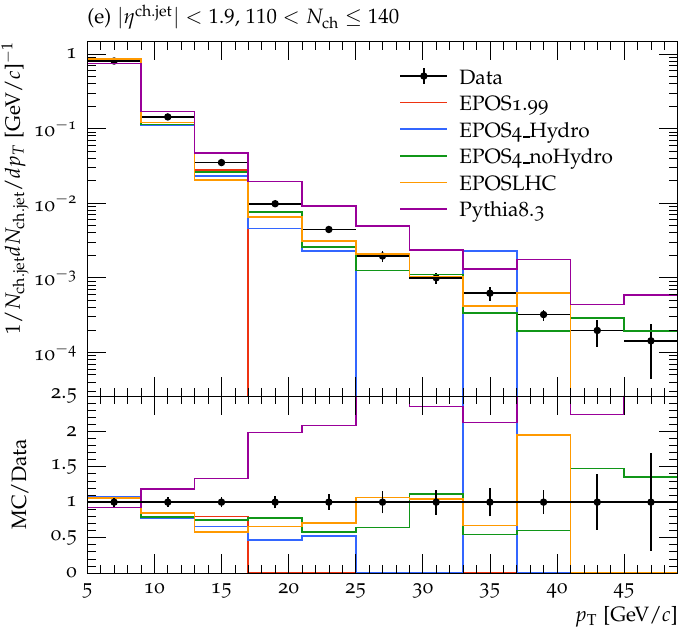}
\caption{ Differential jet $p_{T}$ spectrum for inclusive charged particles in an event in (a) $10<N_{ch}\leq30$ (b)$30<N_{ch}\leq50$ (c) $50<N_{ch}\leq80$ (d) $80<N_{ch}\leq110$ and (e) $110<N_{ch}\leq150$.Result of different Monte Carlo models Pythia8.308\cite{STM}, EPOS1.99\cite{epos1.99}, EPOSLHC\cite{eposL}, EPOS4\cite{epos4a} compared with the experimental data \cite{J}}.\label{Fig3}
\end{center}
\end{figure*}
Figure~\ref{Fig3} illustrates differential jet $p_{T}$ spectrum for inclusive charged particles in an event in (a) $10<N_{ch}\leq30$ (b) $30<N_{ch}\leq50$ (c) $50<N_{ch}\leq80$ (d) $80<N_{ch}\leq110$ and (e) $110<N_{ch}\leq150$. 
When comparing Monte Carlo (MC) models to the data, it is observed that for $10<N_{ch}\leq30$, all the models show good agreement with the data at $p_{T}<10$ GeV/c. The EPOSLHC model tends to overestimate the data for $10<p_{T}<25$ GeV/c and then underestimates for $25<p_{T}<35$ GeV/c. The EPOS4$_{Hydro}$ model accurately reproduces the data for the $5<p_{T}<35$ GeV/c range, while Pythia8 overestimates and EPOS4$_{noHydro}$ underestimates the data for $p_{T}>10$ GeV/c. For $30<N_{ch}\leq50$ and $50<N_{ch}\leq80$, all the models accurately predict the data at $p_{T}<10$ GeV/c. For the $10<p_{T}<35$ GeV/c range, EPOSLHC and EPOS1.99 reproduce the results well, and EPOS4$_{noHydro}$ underestimates, while Pythia8 overestimates with the data. EPOS4$_{Hydro}$ disagrees with the data in the $N_{ch}>15$ region. For $80<N_{ch}\leq110$ and $110<N_{ch}\leq150$, again, all models reproduce the data well for $5<p_{T}<10$ GeV/c. For the intermediate and high $p_{T}$ region, Pythia8 overestimates, while EPOS4$_{Hydro}$ and EPOS4$_{noHydro}$ agree well with the data. EPOS1.99 underestimates the data for the $15<p_{T}<35$ GeV/c region, and for the higher $p_{T}$ region, it fails to reproduce the results.
\begin{figure*}[p!]
\begin{center}
\hskip-0.153cm
\includegraphics[width=8cm]{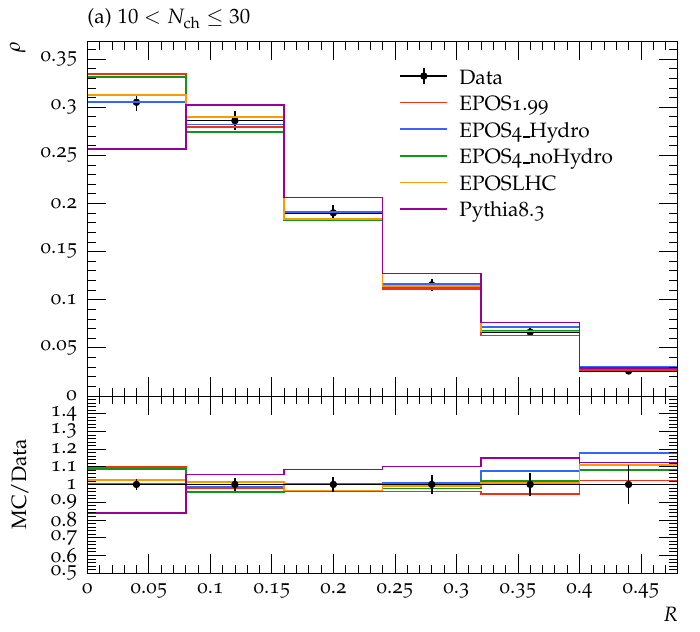}
\includegraphics[width=8cm]{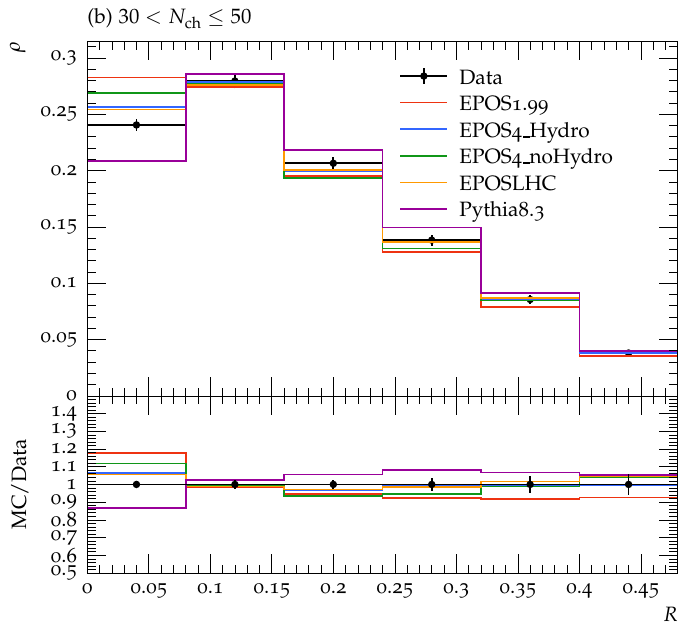}
\includegraphics[width=8cm]{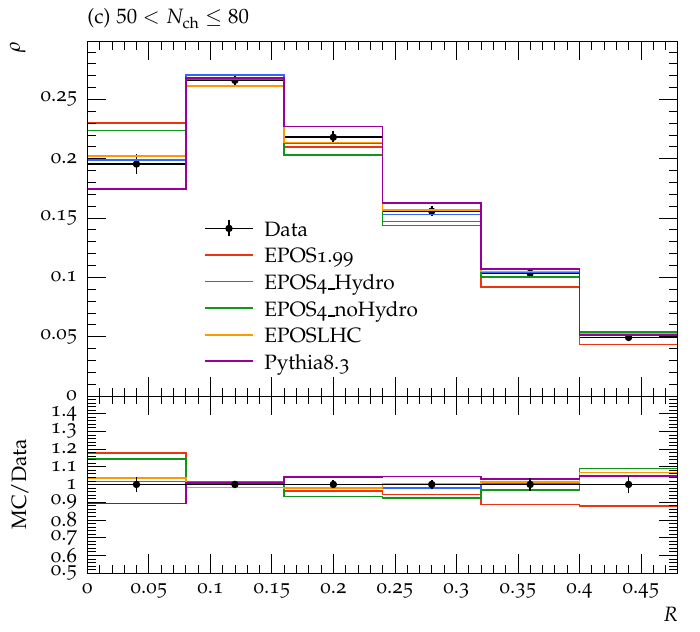}
\includegraphics[width=8cm]{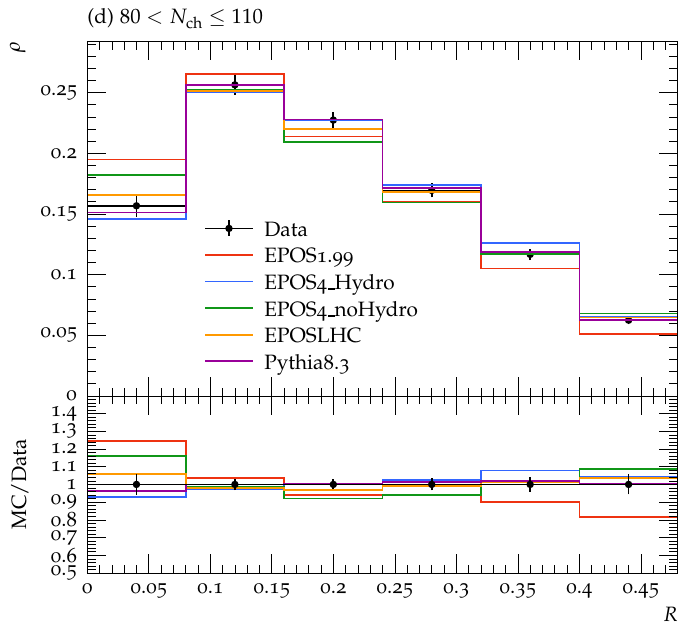}
\includegraphics[width=8cm]{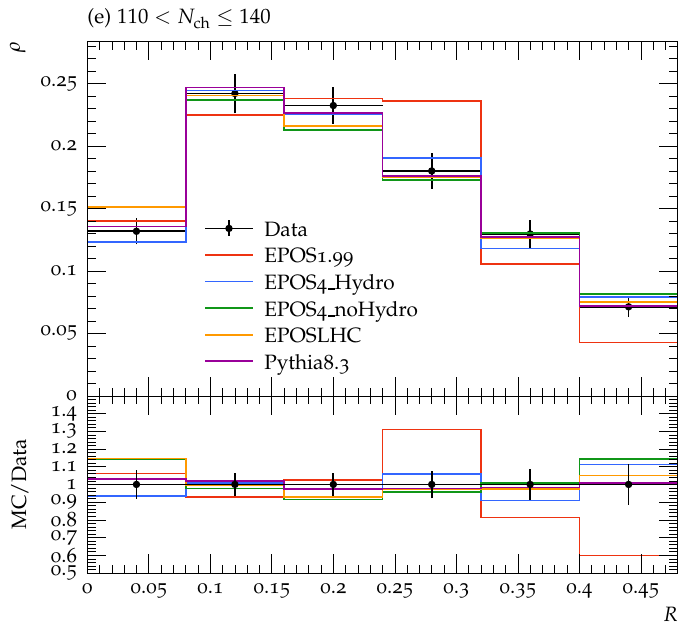}
\end{center}
 \caption {Normalized charged-particle jet $p_{T}$ density $\rho$ as a function of distance
to the jet axis R in (a) $10<N_{ch}\leq30$ (b)$30<N_{ch}\leq50$ (c) $50<N_{ch}\leq80$ (d) $80<N_{ch}\leq110$ and (e) $110<N_{ch}\leq140$. Result of different Monte Carlo models Pythia8.308\cite{STM}, EPOSLHC\cite{eposL}, EPOS1.99\cite{epos1.99}, EPOS4\cite{epos4a} compared with the data \cite{J}}.\label{Fig4}
\end{figure*}
Figure~\ref{Fig4} illustrates the normalized charged-particle jet \( p_{T} \) density as a function of the distance to the jet axis \( R \) for events in five \( N_{ch} \) intervals. The results of the MC predictions are compared with the data. The jet \( \rho \) increases as \( N_{ch} \) increases. For the \( 10 < N_{ch} \leq 30 \) and \( 30 < N_{ch} \leq 50 \) intervals, when different MC predictions are compared with the data, it is observed that for \( R < 0.05 \), EPOS4\(_{Hydro}\) and EPOSLHC provide good predictions, Pythia8 underestimates, and EPOS4\(_{noHydro}\) overestimates the data. For larger distances, the Pythia8 model overestimates the data, while all the remaining models show excellent agreement with the data. EPOS4\(_{Hydro}\) and EPOS1.99 accurately reproduce the data for the entire distance.
For the \( 50 < N_{ch} \leq 80 \) and \( 80 < N_{ch} \leq 110 \) ranges, when \( R < 0.1 \), EPOS4\(_{Hydro}\) and EPOSLHC provide good predictions, while Pythia8 underestimates and EPOS4\(_{noHydro}\) overestimates the data. For the \( 0.1 < R < 0.45 \) region, all these models accurately reproduce the data. EPOS1.99 underestimates the data for higher distances from the jet axis. 
For the \( 110 < N_{ch} \leq 140 \) range, all the above-mentioned models overestimate the data except for EPOS4\(_{Hydro}\). However, in the intermediate region, EPOS4\(_{Hydro}\), EPOS4\(_{noHydro}\), Pythia8, and EPOSLHC provide good predictions, while EPOS1.99 underestimates the data. For larger values of \( R \), EPOS4\(_{noHydro}\), EPOS4\(_{Hydro}\), EPOSLHC, and Pythia8 show good agreement with the data. However, EPOS1.99 performs differently, failing to reproduce the data for the \( R > 0.25 \) range.

\section{Summary and Conclusion}
In this study, we present a comprehensive analysis of jets and underlying events as a function of charged particle multiplicity in proton-proton (pp) collisions at a center-of-mass energy of \(\sqrt{s} = 7\) TeV. Various Monte Carlo (MC) event generators, including Pythia8.308, EPOS1.99, EPOSLHC, EPOS4\(_{Hydro}\), and EPOS4\(_{noHydro}\), are employed to predict particle production. The predictions from these models are compared with experimental data from the CMS collaboration. The produced particles are divided into two classes: those associated with underlying events and those associated with jets. The charged particles are tracked within \(|\eta| < 2.4\) and \(p_{T} > 0.25 \, \text{GeV}/c\), while charged particle jets are calculated with \(p_{T} > 5 \, \text{GeV}/c\) using only charged particle information. In this work, we present jet \(p_{T}\) distributions, the mean \(p_{T}\) of underlying event and jet particles, jet rates, and normalized charged density as a function of \(N_{ch}\). We observed that the mean transverse momentum for all charged particles and underlying event charged particles increases with rising charged-particle multiplicity. This indicates that at higher \(N_{ch}\), multiple parton interactions are increasing, and hard scattering is occurring.
When comparing the Monte Carlo (MC) predictions with the CMS data, we found that all models agree well with the data at low \( N_{ch} \). However, at higher \( N_{ch} \), only EPOS4\(_{Hydro}\) and EPOSLHC generally align with the data. On the other hand, the mean \( p_{T} \) for intra-jet and leading charged particle jets decreases logarithmically as the charged particle multiplicity increases. This indicates an opposite trend: a larger number of multiple parton interactions occur due to softer processes. Consequently, the production of final state hadrons is attributed to (mini)jets. EPOS4\(_{Hydro}\), Pythia8, and EPOSLHC reproduce the results, while EPOS1.99 overestimates and EPOS4\(_{noHydro}\) underestimates the data. For charged particle jets, EPOS4\(_{Hydro}\) and EPOSLHC align with the data. These results suggest that EPOS models with a hydrodynamic component perform best as the charged particle multiplicity increases. This is because "tuning on Hydro" increases the multiplicity, converting some energy into the flow, which allows EPOS4\(_{Hydro}\) to closely match the data and produce accurate results at higher \( N_{ch} \) compared to other models. The similarity between Pythia8 and EPOSLHC models arises from their use of partonic methods and perturbative approaches for describing hard collisions. 
When studying the charged jet rates, we observed that for the \( p^{\text{ch.jet}} > 5 \, \text{GeV}/c \) range, only Pythia8 and EPOS4\(_{Hydro}\) performed the best, with EPOSLHC also fitting the results well. EPOS1.99 initially under-predicted the data and then over-predicted it, while EPOS4\(_{noHydro}\) consistently underestimated the data and did not perform well. For \( p^{\text{ch.jet}} > 30 \, \text{GeV}/c \), no Monte Carlo model performed best at low \( N_{ch} \); however, at higher charged particle multiplicity, only EPOS4\(_{Hydro}\) performed well. This is due to the "Hydro = on" option, which reduces the multiplicity and converts some energy into flow, making the blue curve much stronger than the other curves. The EPOSLHC model also described the data well, owing to its incorporation of new collective flow effects and correlated flow treatment, which contrasts with the EPOS1.99 model.
In the study of differential jet $p_{T}$ spectrum for five intervals, we observed that for the selected intervals all the given MC models good agree with the result at low and intermediate $p_{T}$, however for higher $p_{T}$, EPOSLHC, EPOS4$_{Hydro}$, EPOS4$_{noHydro}$ well agree with the data. Pythia8 overestimates while EPOS1.99 and EPOS4$_{Hydro}$ underestimate the data and do not well reproduce the result for higher $p_{T}$.
In the study of normalized charged $p_{T}$ jet density for five intervals, we observed that only EPOS4$_{Hydro}$ and EPOSLHC perform the best at low $p_{T}$ however, at larger $p_{T}$, all the given MC models good agree with the data. EPOS1.99 under-predict the data as $N_{ch}$ was increasing. EPOSLHC model described the data well due to the new collective flow effects, correlated flow treatment, and parametrization as compared to the EPOS1.99. This suggests that each model has its advantages and limitations, which vary depending on the specific physical scenario. The suitability of a particular model is determined by the context of the investigation.
\\
\\
{\bf Acknowledgments:}
The present research work was funded by Princess Nourah bint Abdulrahman University Researchers Supporting Project number (PNURSP2024R106), Princess Nourah bint Abdulrahman University, Riyadh, Saudi Arabia. 
\\
\\
{\bf Author Contributions:} All authors listed have made a
substantial, direct, and intellectual contribution to the work and
approved it for publication.
\\
\\
{\bf Data Availability Statement:} The data used to support the findings of this study are
included within the article and are cited at relevant places
within the text as references.
\\
\\
{\bf Compliance with Ethical Standards:}
The authors declare that they are in compliance with ethical standards regarding the content of this
paper.
\\
\\
\\
\\
{\bf Conflict of Interest:} The authors declare that there are no conflicts of interest regarding the publication of this paper.
\\
\\

\end{document}